\documentclass[10pt,a4paper,onecolumn]{article}
\usepackage{graphicx}
\usepackage{color}
\begin{document}
\title{Heat engine model exhibit super-universal feature and 
capture the efficiencies of different power plants}

\author{M. Ponmurugan \\
Department of Physics, School of Basic and Applied Sciences, \\
Central University of Tamilnadu, Thiruvarur - 610 005,  \\
Tamilnadu, India. e-mail:ponphy@cutn.ac.in}

%\author{M. Ponmurugan}
%\email[]{ponphy@cutn.ac.in}
%\affiliation{Department of Physics, School of Basic and Applied Sciences, Central
%University of TamilNadu, Thiruvarur 610 005, TamilNadu, India.}
%\date{\today}

%\pacs{05.70.Ln}{Non-equilibrium and irreversible thermodynamics}
%\pacs{05.70.-a}{Thermodynamics}
%\pacs{05.20.-y}{Classical statistical mechanics}
%\Keyword:}  Heat engine, irreversible thermodynamics, efficiency, universality

\maketitle

\begin{abstract}
We propose a generalized model of a heat engine and calculate the minimum and maximum  
bounds on the efficiency at maximum power. We obtain a universal form of generalized extreme bounds 
on the efficiency at maximum power. Our model unifies the bounds on 
the efficiency and the universality features are observed for various heat engine models. 
Even though our model is a direct generalization of
low-dissipation heat engines, the bounds on the efficiency obtained at a single 
target function capture those observed in the actual power plants
working at different dissipation levels.
\end{abstract}

\section{Introduction}

	In the study of non-equilibrium thermodynamics, the universal features of the heat engine efficiency 
at maximum target function have attracted  much attention in the past few years 
\cite{targetfn1,targetfn2,targetfn3}. Heat engine is a thermodynamic system operating between hot and cold heat reservoirs. The engine performs useful work $W$ by absorbing $Q_h$ amount of heat from the hot reservoir at a higher temperature $T_h$ and delivers $Q_c$ amount of heat to the cold reservoir at a lower temperature $T_c$. The efficiency of the heat engine operating 
between two reservoirs is defined as $\eta=W/Q_h$, which is bounded below the Carnot efficiency, $\eta_C=1-T_c/T_h$. 

In equilibrium thermodynamics, the Carnot efficiency is the maximum efficiency reached by the heat engine when
it is operating in the quasi-static process of infinitely long time duration in which the power delivered by the engine is zero. 
The quasi-static process can also be realized in a finite time for micro-sized heat engines kept in  
optical or bacterial reservoirs and operated at a single molecule level \cite{nat1,nat2,holuprl}. 
However, the actual power plants are macroscopic engines which are operating in the 
regime of near equilibrium or completely in the non-equilibrium regime. 
The efficiency observed by the macroscopic heat engines should be less than $\eta_C$ since the engines are operating in the finite time duration of non-zero power output.

Most of finite time thermodynamics studies focus on  maximizing different target 
functions \cite{targetfn1,targetfn2,targetfn3} to analyze the heat engine performances 
in the linear and non-linear regimes \cite{book1}.
The most frequently used target functions are the efficiency 
at maximum power, $\eta_P$ \cite{targetfn1,Espo}, the efficiency at maximum figure of merit or efficient power,
 $\eta_{\chi}$ (a product of $\eta$ and power $P$) \cite{effpr,holubec1,lowchi1} and the efficiency at maximum $\Omega$ (ecological criterion), $\eta_{\Omega}$, which accounts both for
useful energy and losses \cite{targetfn3,eco,lowOmega}.
 
The salient feature of heat engine efficiency at maximum target function is that one can obtain the 
universal form of efficiency up to  quadratic order in $\eta_C$, irrespective of the different models 
such as stochastic heat engine \cite{stocheng}, low-dissipation heat engine \cite{Espo},
minimally nonlinear heat engine \cite{nonlinheat}, etc. In other words,  the efficiency at maximum 
target function displays the universality up to quadratic order when the system deviates from equilibrium. 
For different target functions, different upper and lower bounds on the efficiency at maximum 
target function and  different universality classes of the efficiency up to quadratic order in $\eta_C$
can be obtained.

For the efficiency at maximum power, the upper and lower bounds on the efficiency are given by
$\frac{\eta_C}{2} \le \eta_P \le \frac{\eta_C}{2-\eta_C}$ and the universal form of the efficiency up to quadratic order in $\eta_C$ is \cite{Espo},
\begin{eqnarray}\label{maxeffuni1} 
  \eta_P &=&\frac{1}{2}\eta_C+\frac{1}{8}\eta_C^2+ O(\eta_C^3).
\end{eqnarray} 
The efficiency at maximum power of an optimized thermal engine in the endoreversible limit 
given by  $\eta_{CA}=1-\sqrt\frac{T_c}{T_h}=1-\sqrt{1-\eta_C}$ \cite{novikov,curzon} is usually 
called as the Curzon-Ahlborn efficiency.  When the temperature difference between the two reservoirs is small, 
the Taylor expansion of $\eta_{CA}$ gives Eq.(\ref{maxeffuni1}) \cite{uniTU,endoqm}, which
is bounded below the Carnot efficiency of the reversible heat engines. However, this model does not 
represent a universal bound on the efficiency at maximum power \cite{book1,Hoffmann}.

Whereas, in the case of efficient power, the upper and lower bounds on the efficiency are given by
$\frac{2}{3} \eta_C \le \eta_{\chi} \le \frac{3-\sqrt{9-8\eta_C}}{2}$ and the universal form of the efficiency up to quadratic order in $\eta_C$ is \cite{lowchi1},
\begin{eqnarray}\label{maxeffuni2}
  \eta_{\chi} &=&\frac{2}{3}\eta_C+\frac{2}{27}\eta_C^2+ O(\eta_C^3)
\end{eqnarray}
and in the case of maximum $\Omega$ criterion, the upper and lower bounds on the efficiency are given by
$\frac{3}{4} \eta_C \le \eta_{\Omega} \le \frac{3-2\eta_C}{4-3\eta_C}\eta_C$ and the universal form of the efficiency up to quadratic order in $\eta_C$ is \cite{targetfn3},
\begin{eqnarray}\label{maxeffuni3}
  \eta_{\Omega} &=&\frac{3}{4}\eta_C+\frac{1}{32}\eta_C^2+ O(\eta_C^3).
\end{eqnarray}
The above results are obtained for different heat engine models under the assumption 
that $T_c$ is very close to $T_h$. Further, in the study of a minimal model of information (I) based heat engine \cite{infoengine}, another universality class for efficiency at maximum power up to quadratic order in $\eta_C$ has been observed: 
\begin{eqnarray}\label{maxeffinfo}
  \eta_{PI} &=&\frac{1}{2}\eta_C+\frac{1}{12}\eta_C^2+ O(\eta_C^3).
\end{eqnarray}

Recent studies on heat engines showed the break down of such universality in the case of the efficiency at maximum power \cite{nonuniv,uniloss}. In particular, a recent study on a quantum dot engine reveals that the universal and non-universal form of efficiency at maximum target function depends on the imposed constraint on the control parameter of the heat engine \cite{nonuniv}. Few other heat engine studies showed the importance of constraint relations between the efficiency and power and  the optimal finite time protocol
to achieve the efficiency at maximum power \cite{modylow1,modylow2}.

The simplest model, which is believed to capture the features of various physical systems is, 
the low-dissipation Carnot engine \cite{Espo}. There are extensive studies on low-dissipation 
Carnot engine model and its performance under maximizing different target functions 
\cite{Espo,lowchi1,lowOmega}. Some of the industrial power plants \cite{Espo,holubec1,lowOmega,Johal}
does not operate in the low-dissipation regime \cite{geneff}. Our proposed model, which is a  direct generalization of the low-dissipation model, is found to be valid for any dissipation levels in which 
the heat engines might operate. 
It also unifies the different universal form of efficiency at 
a single target function (maximum power). Even though the operation regimes are completely different for 
different target functions, the minimum and maximum bounds on the efficiency 
at maximum power for different dissipation levels encompass all the extreme bounds on the efficiency at 
different target functions studied so far.
Below, we will briefly summarize the existing low-dissipation model and give a  
detailed explanation of  our generalized model.

\section {Low-dissipation model}
The considered low-dissipation model describes a heat engine undergoing Carnot cycle, consisting  of two isothermal processes 
of finite time duration and two instantaneous adiabatic processes. In the isothermal expansion (compression),
the  working substance is in contact with the hot (cold) reservoir at temperature $T_h$ ($T_c$) during the time interval  $t_h$($t_c$). The amount of heat $Q_h$ ($Q_c$) exchanged between the hot (cold) reservoir and the working substance is modeled as \cite{Espo}
\begin{eqnarray}\label{qh}
 Q_h=T_h \left( \Delta S -\Sigma_h / t_h \right),  
\end{eqnarray}
\begin{eqnarray}\label{qc}
 Q_c=T_c \left(-\Delta S - \Sigma_c / t_c \right),  
\end{eqnarray}
where $\Sigma_h$, $\Sigma_c$ are the dissipation coefficients, describing the irreversibility present in the model, and $\pm \Delta S$ is the change in entropy of the working substance during  isothermal expansion (+) and compression (-), such that the total change in entropy of the working substance is zero \cite{lowchi1}.

The work performed by the engine during  the  total time period $t=t_h+t_c$ is $-W=Q_h+Q_c$. Here, we used 
the convention that work and heat absorbed by the system are positive \cite{Espo}. 
The power ($P$) generated during the Carnot cycle is given by, 
\begin{eqnarray}\label{pow}
  P =\frac{-W}{t_h+t_c}=\frac{(T_h-T_c) \Delta S - T_h\Sigma_h / t_h - T_c \Sigma_c / t_c } {t_h+t_c}. 
\end{eqnarray}
Using Eq.(\ref{qh}), $Q_c$ can be rewritten as 
 \begin{eqnarray}\label{qcmod}
  Q_c=-T_c \left(Q_h/T_h+\Sigma_h/t_h+\Sigma_c/t_c \right).  
\end{eqnarray}
Then the engine efficiency during the Carnot cycle is given by  
 \begin{eqnarray}\label{effg}
  \eta =\frac{Q_h+Q_c}{Q_h} =\eta_C - T_c \left(\frac{\Sigma_h/t_h+\Sigma_c/t_c}{Q_h}\right).  
\end{eqnarray}
From the above equation, it is observed that the relation between the efficiency and the entropy production per engine cycle is similar to what is stated in Ref.\cite{Lee1}. Here, 
%\begin{eqnarray}\label{qcomb}
  $Q_h+Q_c  = \eta_C Q_h- T_c \left(\Sigma_h/t_h+\Sigma_c/t_c \right)$.
%\end{eqnarray}
Using Eq.(\ref{qh}), Eq.(\ref{effg}) can be rewritten as 
\begin{eqnarray}\label{effg1}
  \eta &=& \eta_C -\frac{T_c}{T_h} \left(\frac{1+\frac{\Sigma_c t_h}{\Sigma_ht_c}}{\frac{ \Delta S t_h}{ \Sigma_h} -1}\right).  
\end{eqnarray}
The values of $t_h$ and $t_c$, at which the power is maximum, as given by \cite{Espo}, are,
$t_h = 2\frac{T_h\Sigma_h}{(T_h-T_c) \Delta S} \left(1+\sqrt{\frac{T_c \Sigma_c}{T_h\Sigma_h}}\right)$
and 
$	t_c = 2\frac{T_c\Sigma_c}{(T_h-T_c) \Delta S} \left(1+\sqrt{\frac{T_h \Sigma_h}{T_c\Sigma_c}}\right)$.
Their ratio obeys the relation, $\frac{t_c}{t_h} = \sqrt{\frac{T_c \Sigma_c}{T_h\Sigma_h}}$.
By using the above equations, the efficiency at maximum power can be obtained and is given by 
\begin{eqnarray}\label{effmaxp}
   \eta_P&=&\eta_C - \frac{\eta_C}{2}\left[\frac{ 1-\frac{\eta_C}{\left(1+\sqrt{\frac{T_c \Sigma_c}{T_h\Sigma_h}}\right)}}{1-\frac{\eta_C}{2\left(1+\sqrt{\frac{T_c \Sigma_c}{T_h\Sigma_h}}\right)}} \right].
\end{eqnarray}
The above equation showed that $\eta_P$, in general, does not exhibit any universal form \cite{nonuniv}.
However, in the asymmetric dissipation limits, $\Sigma_c/\Sigma_h \to \infty$ and  $\Sigma_c/\Sigma_h \to 0$,
$\eta_P$ converges respectively to the lower bound $\eta_C/2$ and the upper bound $\eta_C/(2-\eta_C)$. 
In the symmetric dissipation, $\Sigma_c=\Sigma_h$, the above equation can be expanded in terms of $\eta_C$ 
as, $\eta_P	= \frac{\eta_C}{2} + \frac{\eta_C^2}{8} + O(\eta_C^3)$.
This universal form of the efficiency at maximum power up to quadratic order in $\eta_C$ is thus obtained  
only under the assumption that the temperature difference between $T_c$ and $T_h$ is small \cite{Espo}.

The inverse proportionality relation between the irreversible entropy production and 
time for the low-dissipation model may be a reasonable assumption for many systems operating in finite time.
Different control schemes are utilized to tune the efficiency at the maximum power
of the low-dissipation heat engine models \cite{holubec1,stocheng,ExpEMP}. 
Recent study also showed that the efficiency at the  maximum power 
of a heat engine can be optimized by tuning the system's energy levels \cite{modylow1}. 
This can be done in  low-dissipation heat engine model by incorporating the control scheme 
on the coefficients $\Sigma_{h,c}$ in the isothermal process \cite{stocheng,modylow1,modylow2}. 
Underlying this control scheme and using the low-dissipation assumption as a base model, 
three different types (normal, sub and super) of dissipative heat engines and 
its bounds were proposed earlier \cite{yang}. However, the significance of 
implementing such a control scheme have not yet been discussed more elaborately 
in general for heat engines operating in different dissipation regimes \cite{yang}.
Therefore, it is worthful to investigate the different dissipation behaviors of such heat 
engines \cite{geneff, yang} in a generalized case.

\section{Generalized model}

Even though the low-dissipation model is a well-founded model for many heat 
engines \cite{holubec1,stocheng}, it might not be suitable for real
heat engines \cite{holubec1,geneff,yang}.  Below,
following Ref.\cite{yang},  we  generalize the low-dissipation model 
of the heat engine and calculate the minimum  and maximum bounds on the efficiency 
at maximum power.

The amount of heat $Q_h$ ($Q_c$) exchanged between the hot (cold) reservoir and the working substance is modeled as
\begin{eqnarray}\label{gqh}
 Q_h=T_h \left\{ \Delta S - \alpha_h \left(\frac{\sigma_h}{t_h}\right)^{\frac{1}{\delta}} \right\},  
\end{eqnarray}
\begin{eqnarray}\label{gqc}
 Q_c=T_c \left\{-\Delta S - \alpha_c \left(\frac{\sigma_c}{t_c}\right)^{\frac{1}{\delta}} \right\},  
\end{eqnarray}
where $\sigma_h=\lambda_h\Sigma_h$, $\sigma_c=\lambda_c\Sigma_c$, $\delta \ge 0$ is a  real number, which 
represents the level of dissipation, and $\lambda_{h/c}$ \& $\alpha_{h/c}$ are the dimensionality-preserving tuning parameters such that the quantity inside the parenthesis (possibly having a nonlinear power $1/\delta$) is dimensionless. 
Note that the parameter $\delta$ departs this model from the first approximation in the entropy 
generation of irreversible heat devices. Under the assumption that $\sigma_{h/c} < t_{h/c}$,
$\delta \to 0$ implies no dissipation, $\delta =1$
low-dissipation and $\delta \to \infty$ high dissipation.

In recent years, the heat engine studies are primarily focused on mesoscopic systems, which require more sophisticated microscopic models \cite{micro, Cavina}. Using the Markovian master equation approach, the study of generalized framework of quantum mechanical heat engines provided the 
solid microscopical basis to the low-dissipation theory \cite{Cavina}.
The phenomenological approach used in our work makes a general impression that some previously derived performance bounds can be recovered, since the employed model contains three additional fitting parameters compared to the standard low-dissipation model. The main physical motivation of our proposed model is that the coefficient $\delta$, which is not necessarily an integer, might capture a non-Markovian 
dynamics of the system \cite{Cavina}.
Similarly to Ref.\cite{yang}, $\lambda_{h/c}$ are related with some external controlled parameter that drives the system during the isothermal processes in a given time interval. Further, our model has an
additional parameter $\alpha_{h/c}$, which might be  related to the control scheme that tune 
the system energy levels during
the isothermal processes \cite{modylow1}. These tuning parameters allow to control the
irreversible entropy generation by choosing a suitable  combination of control schemes.
Using Eq.(\ref{gqh}), $Q_c$ can be rewritten as 
 \begin{eqnarray}\label{gqcmod}
  Q_c=-T_c \left[Q_h/T_h+\alpha_h(\sigma_h/t_h)^\frac{1}{\delta}+ \alpha_c(\sigma_c/t_c)^\frac{1}{\delta} \right].  
\end{eqnarray}
Then,
\begin{eqnarray}\label{gqcomb}
  Q_h+Q_c = \eta_C Q_h- T_c \left[\alpha_h(\sigma_h/t_h)^\frac{1}{\delta}+\alpha_c(\sigma_c/t_c)^\frac{1}{\delta} \right].
\end{eqnarray}
The engine efficiency for the Carnot cycle is  
 \begin{eqnarray}\label{geffg}
  \eta =\eta_C - T_c \left[\frac{\alpha_h(\sigma_h/t_h)^\frac{1}{\delta}+\alpha_c(\sigma_c/t_c)^\frac{1}{\delta}}{Q_h}\right].  
\end{eqnarray}
As mentioned above, this equation, relating the efficiency and the entropy production per engine 
cycle, is similar to the relation given in Ref.\cite{Lee1}. Further, 
using Eq.(\ref{gqh}), the above equation can be rewritten as 
\begin{eqnarray}\label{geffg1}
  \eta &=& \eta_C -\frac{T_c}{T_h} \left(\frac{1+\frac{\alpha_c}{\alpha_h}\left(\frac{\sigma_c t_h}{\sigma_h t_c}\right)^\frac{1}{\delta}}{ \frac{\Delta S}{\alpha_h} \left(\frac{t_h}{\sigma_h}\right)^\frac{1}{\delta} -1}\right)  
\end{eqnarray}
and the power generated during the Carnot cycle is given by 
\begin{eqnarray}\label{gpow}
  P &=& \frac{(T_h-T_c) \Delta S - T_h \alpha_h(\sigma_h / t_h)^\frac{1}{\delta} - T_c \alpha_c(\sigma_c / t_c)^\frac{1}{\delta}} {t_h+t_c}.
\end{eqnarray}
The values of $t_h$ and $t_c$ at which the power becomes maximum are given by,
%\begin{widetext}
\begin{eqnarray}\label{gtopt}
  t_h &=&\left \{ \left(1+\frac{1}{\delta}\right)\alpha_h\frac{T_h\sigma_h^\frac{1}{\delta}}{(T_h-T_c) \Delta S} \left[1+\left(\frac{\alpha_c T_c}{\alpha_h T_h}\right)^\frac{\delta}{\delta+1}\left(\frac{ \sigma_c}{\sigma_h}\right)^\frac{1}{\delta+1} \right] \right \}^{\delta}, \\
	t_c &=& \left \{ \left(1+\frac{1}{\delta}\right)\alpha_c\frac{T_c\sigma_c^\frac{1}{\delta}}{(T_h-T_c) \Delta S} \left[1+\left(\frac{\alpha_h T_h}{\alpha_c T_c}\right)^\frac{\delta}{\delta+1}\left(\frac{\sigma_h}{\sigma_c}\right)^\frac{1}{\delta+1} \right] \right \}^{\delta},
\end{eqnarray} 
%\end{widetext}
and the ratio between the two obeys the relation
\begin{eqnarray}\label{gtratio}
 \left(\frac{t_c}{t_h}\right)^{\frac{1}{\delta}+1} = \frac{\alpha_c T_c}{\alpha_h T_h}\left(\frac{\sigma_c}{\sigma_h}\right)^\frac{1}{\delta}.
\end{eqnarray}
By using Eqs.(\ref{geffg1}), (\ref{gtopt}) and (\ref{gtratio}), the efficiency at maximum
 power can be obtained as, 
\begin{eqnarray}\label{geffmaxp}
   \eta_P	&=& \left(\frac{1}{\delta+1}\right)\frac{\eta_C}{1-\frac{\eta_C}{(1+\frac{1}{\delta})\zeta}},
\end{eqnarray}
where, 
\begin{eqnarray}\label{chie}
\zeta=1+\varsigma \left(\frac{T_c}{T_h}\right)^\frac{\delta}{\delta+1}=1+\varsigma \left(1-\eta_C\right)^\frac{\delta}{\delta+1} 
\end{eqnarray}
and 
\begin{eqnarray*}\label{tun}
\varsigma&=&\left(\frac{\alpha_c}{\alpha_h}\right)^\frac{\delta}{\delta+1}\left(\frac{\sigma_c}{\sigma_h}\right)^\frac{1}{\delta+1}.
\end{eqnarray*}
Note that the efficiency at maximum power does not depend on the individual parameters but only on their ratios $\sigma_c/\sigma_h$ and $\alpha_c/\alpha_h$.
The above equation  shows that $\eta_P$  in general does not exhibit any universal form. However, in the asymmetric dissipation limits, $\sigma_c/\sigma_h \to \infty$ or $\zeta \to \infty$
and  $\sigma_c/\sigma_h \to 0$ or $\zeta \to 1$,
$\eta_P$ converges respectively to its lower bound $\eta^{-}_P=\frac{1}{\delta+1}\eta_C$ and its upper bound 
$\eta^{+}_P=\frac{\eta_C}{(\delta+1)-\delta \eta_C}$ and thus it is bounded as
\begin{eqnarray}\label{gbound}
\frac{1}{\delta+1}\eta_C \le \eta_P  \le  \frac{\eta_C}{(\delta+1)-\delta \eta_C}. 
\end{eqnarray}
This result was obtained earlier in Ref.\cite{yang} by using  
power law profile control scheme. Nevertheless, the significance of the above 
relation have not yet discussed more elaborately. In what follows, we will discuss the 
universal form of the above relation for different dissipation levels in detail.

When $\delta=1$, one can recover the results of low-dissipation heat engines \cite{Espo} 
with $\alpha_c=\alpha_h=1$ and $\lambda_c=\lambda_h=1$. 
As mentioned earlier, the operation regimes are entirely different for different target
functions. Nevertheless, we find that the minimum and maximum bounds on the efficiency
at maximum power span the extreme bounds on the efficiency of different
target functions with a proper choice of $\delta$.

For $\delta=1/2$, one gets $\eta^{-}_P=(2/3)\eta_C$ and $\eta^{+}_P=2\eta_C/(3-\eta_C)$, 
we obtain the same lower bound on the efficiency at maximum  efficient power of low-dissipation Carnot like engines and other models of nonlinear irreversible heat engines \cite{lowchi1,multitarget}.  
Similarly, for $\delta=1/3$, $\eta^{-}_P=(3/4)\eta_C$  and $\eta^{+}_P=3\eta_C/(4-\eta_C)$, 
we obtain the same lower bound on the efficiency at maximum  $\Omega$ criterion for
low-dissipation Carnot like engines and  different  models of linear and 
nonlinear irreversible heat engines \cite{lowOmega,multitarget}. 
However, the upper bounds calculated here are different from (in fact, higher than) those 
generally observed for other target functions.

The generalized minimum and maximum bounds on the efficiency at a single target function 
(maximum power) was obtained that covers the extreme bounds obtained for other target functions  
\cite{lowchi1,lowOmega,multitarget}. It should  be emphasized that in our generalized model, 
the extreme bounds on the efficiency at maximum power for $\delta < 1$ clearly indicate 
that the heat engines dissipate  in those regimes less than in the low-dissipation regime.

%%%%%%%%%%%%%%%%%%%%%%%%%%%%%%%%%%%%%%%%%%%%%%%%%%%%%%%%%%%%
\begin{table}
\centering
\caption{
Expansion of the efficiency at maximum power
$\eta_P=\left(\frac{1}{\delta+1}\right)\eta_C +\frac{\delta}{(\delta+1)^2 (1+\varsigma)}\eta_C^2+O(\eta_C^3)$ 
up to quadratic order in $\eta_C$ for $\varsigma=1$ to $5$
and for three dissipation levels $\delta$.  
}
\label{tab:univ} 
\begin{tabular}{|c|c|c|c|}
 \hline
&&& \\ 
$\varsigma$     &   $\delta=1$     &  $\delta=1/2$   &  $\delta=1/3$ \\  
&&& \\ \hline    
&&& \\ 
 1  & $\frac{1}{2}\eta_C+\frac{1}{8}\eta_C^2$ & $\frac{2}{3}\eta_C+\frac{1}{9}\eta_C^2$ &  $\frac{3}{4}\eta_C+\frac{3}{32}\eta_C^2$ \\ 
&&& \\  
2  & $\frac{1}{2}\eta_C+\frac{1}{12}\eta_C^2$ & $\frac{2}{3}\eta_C+\frac{2}{27}\eta_C^2$ &  $\frac{3}{4}\eta_C+\frac{1}{16}\eta_C^2$  \\ 
&&& \\ 
 3  & $\frac{1}{2}\eta_C+\frac{1}{16}\eta_C^2$ & $\frac{2}{3}\eta_C+\frac{1}{18}\eta_C^2$ &  $\frac{3}{4}\eta_C+\frac{3}{64}\eta_C^2$  \\ 
&&& \\ 
 4  & $\frac{1}{2}\eta_C+\frac{1}{20}\eta_C^2$ & $\frac{2}{3}\eta_C+\frac{2}{45}\eta_C^2$ &  $\frac{3}{4}\eta_C+\frac{3}{80}\eta_C^2$  \\ 
&&& \\  
5  &$\frac{1}{2}\eta_C+\frac{1}{24}\eta_C^2$ &$\frac{2}{3}\eta_C+\frac{1}{27}\eta_C^2$ &  $\frac{3}{4}\eta_C+\frac{1}{32}\eta_C^2$ \\ %\\\hline 
&&& \\\hline  
\end{tabular}
\end{table}
%%%%%%%%%%%%%%%%%%%%%%%%%%%%%%%%%%%%%%%%%%%%%%%%%%%%%%%%%%%%

In the symmetric dissipation $\sigma_c=\sigma_h$, the efficiency at maximum power  
(Eq.\ref{geffmaxp}) becomes,
\begin{eqnarray}\label{geffmaxpsym}
   \eta^{s}_P &=& \frac{\eta_C}{(\delta+1)-\frac{\delta\eta_C}{\zeta_{s}}},
\end{eqnarray}
where 
$\zeta_{s}=1+\left(\frac{\alpha_c T_c}{\alpha_h T_h}\right)^\frac{\delta}{\delta+1}$.
From the above relation, another interesting generalized expression for efficiency at maximum power
is also obtained. Under the tuning condition, $\alpha_c/\alpha_h=T_h/T_c$, and $\zeta_{s}=2$, 
Eq.(\ref{geffmaxpsym}) becomes
 \begin{eqnarray}\label{geffmaxpsym1}
   \eta^{s}_P &=& \frac{\eta_C}{(\delta+1)-\frac{\delta\eta_C}{2}}.
\end{eqnarray}
In the low-dissipation level of $\delta =1$, the above equation further
reduces to $\frac{\eta_C}{2-\frac{\eta_C}{2}}$,
as obtained earlier in the stochastic heat engine model \cite{stocheng}.
These results show that our generalized model comprises the universal expression of the efficiency 
at maximum power for various heat engine models. In the following section, we will investigate 
the universal form of  the efficiency at maximum power obtained from the proposed model.

\section{Universal form}
In order to find  the universal form of the optimized efficiency, we expand Eq.(\ref{geffmaxp})  
in terms of $\eta_C$ as,
\begin{eqnarray}\label{finalexp}
  \eta_P=\left(\frac{1}{\delta+1}\right)\eta_C +\frac{\delta}{(\delta+1)^2 (1+\varsigma)}\eta_C^2+
	\frac{\delta^2}{(\delta+1)^3 (1+\varsigma)^2}\eta_C^3+.....
\end{eqnarray}
The above result shows that the generalized model does not exhibit a universal form 
of the efficiency at maximum power in general. However, it shows the universal 
form for some specific conditions, namely for $\sigma_c/\sigma_h \to 0$,
$\varsigma \to 0$ and $\sigma_c/\sigma_h \to \infty$, $\varsigma \to \infty$. 
The expansions of $\eta_P$ for different values of $\varsigma$ and $\delta$ is given in Table.\ref{tab:univ}.

%%%%%%%%%%%%%%%%%%%%%%%%%%%%%%%%%%%%%%%%%%%%%%%%%%%%%%%%%%%
\begin{table}
\centering
\caption{
The minimum ($\eta^-_P$) and the maximum ($\eta^+_P$) bounds on the efficiency 
at maximum power capture the observed efficiency ($\eta_o$) of industrial
power plants for different dissipation levels $\delta$. $\eta_C$ is the 
Carnot efficiency.
}
\label{tab:PP1} 
\begin{tabular}{|c|c|c|c|c|c| p{1.25in}|}
 \hline 
   Thermal plant   &   $\eta_C$     &   $\eta_o$   &   $\delta$  & $\eta^-_P$ &  $\eta^+_P$ \\ 
  
%& & & & &   \\ 
\hline
%& & & & &   \\ 
CANDU, (Nuclear, Canada)      & 0.48     &   0.30  &  0.6 & 0.30 &  0.36         \\

Calder Hall, (Nuclear, UK)       & 0.49     &   0.19  &  1.6  & 0.19 &  0.27       \\

Steam, UK       & 0.57     &   0.28  &  1.04  &  0.28  &  0.39    \\ 
 
Gas turbine, (Switzerland) & 0.69    & 0.32  &  1.16  & 0.32 & 0.51 \\ \hline  
\end{tabular}
\end{table}
%%%%%%%%%%%%%%%%%%%%%%%%%%%%%%%%%%%%%%%%%%%%%%%%%%%%%%%%%%%%

For $\delta=1$, the universal  form of efficiency at maximum power up to quadratic order
(Eq.(\ref{maxeffuni1})) is  obtained for $\varsigma=1$.  
The present model also captures the  universal form of efficiency at maximum power 
up to quadratic order (Eq.(\ref{maxeffinfo})) for $\varsigma=2$ at the same value of $\delta$
for the minimal model of information based heat engine \cite{infoengine}.
For $\delta < 1$, the universal form of efficiency 
for different target functions (Eqs.(\ref{maxeffuni2}) and (\ref{maxeffuni3})) holds for 
different values of $\varsigma$ as seen in Table.\ref{tab:univ}.
That is, for different dissipation levels, $\delta=1/2$ and $1/3$, 
other universal form of efficiency at maximum 
power can be obtained which are similar to those for the efficient power, multi-parameter 
target functions \cite{multitarget} and $\Omega$, or ecological criterion, 
respectively, for $\varsigma=2, 3,$ and $5$.
Since our model unifies the different universal form of the efficiency at a single target function 
(maximum power) for different dissipation regimes, we can say that Eq.(\ref{finalexp}) exhibits 
a super-universal feature.

\section{Efficiency of industrial power plants}
In order to see whether the generalized model  captures  efficiencies of the 
various industrial power plants, we compared the corresponding observed efficiencies $\eta_o$ 
\cite{Espo,holubec1,lowOmega,Johal,geneff} with the extreme bounds (\ref{gbound}) on the efficiency 
at maximum power. 
The efficiencies of some  power plants are actually captured by the 
extreme bounds for the low-dissipation heat engine corresponding to different target 
functions \cite{Espo,holubec1,lowOmega}. However,
$\eta_o$ of few power plants (UK: Calder Hall, Steam and Switzerland: Gas turbine)
does not fall  within these extreme bounds. 
This indicates that the above power plants are either not operated in the low-dissipation regime
or they are not optimized with respect to the considered target functions.

The extreme generalized bounds on efficiency at maximum power (\ref{gbound}),  
capture the observed efficiencies of actual thermal plants with different values of $\delta$ 
given in Table \ref{tab:PP1}. The value of $\delta$ at which 
 the minimum bound $\eta^-_P=\eta_o$  is calculated  is given by $\delta=\frac{\eta_C}{\eta_o}-1$.
From Table \ref{tab:PP1}, it is clear that $\delta$ larger than one is observed 
for three different power plants. This indicates that these engines might operate
in the region with higher dissipation than the low-dissipation regime $(\delta=1)$. Whereas, 
for $\delta$ less than one, the engines may operate in a regime with smaller  dissipation
than the low-dissipation regime.

The most important factor affecting the parameter $\delta$ for different power plants 
is the way in which the machines exchange heat with the surroundings \cite{geneff}. 
In the case of nuclear plants, the coolant inlet/outlet temperature and pressure
also play a significant role in the machine's performance \cite{iee}. For example, in Table \ref{tab:PP1},  
the steam generating heavy water reactor in UK is similar to the pressurized heavy water reactor 
CANDU in Canada. Heavy water is used as a moderator for both the reactors and 
coolant for CANDU, while ordinary light water is used as a coolant for Steam \cite{iee}. 
Further, the operating conditions 
namely, coolant inlet/outlet temperature and pressure are also different for both  reactors. 
For Calder Hall Magnox reactor in  UK,  graphite is used as a moderator and  carbon dioxide is 
used as a coolant for heat transfer \cite{various1,various2}. Hot gas converts water to steam 
in a steam generator and four heat exchangers generate high and low pressure steam at the same time.
Apart from changing the cladding in the Magnox reactor, the cooling gas pressure also needs to be 
increased for better performance. Finally, the Gas turbine in Switzerland is different from the 
nuclear reactors. The highly efficient axial compressor used in the gas turbine absorbs nearly  
seventy percent of the power from an air inlet at a particular temperature. The heat of the exhaust gases 
is utilized for the production of steam in exhaust boiler which in turn operates with 
low efficiency and high dissipation \cite{geneff}. However, the cycle efficiency  
increases  with the increase in turbine inlet temperature.

\section{Conclusion}
	We generalized the low-dissipation model of a heat engine and obtained 
	the minimum and maximum bounds on the efficiency at maximum power. 
	Extreme bounds on the efficiency at a single target 
function of maximum power capture the efficiency at the maximum power of various heat engine 
models and also efficiencies corresponding to other target functions.

Also the bounds on the efficiency obtained in our generalized model capture the efficiency 
observed in the actual power plants. In  the high dissipation case of 
$\delta \to \infty$, the efficiency $\eta_P$ at maximum power (\ref{geffmaxp}) vanishes,
and for no dissipation, $\delta \to 0$, $\eta_P \to \eta_C$.  
The study of  attainability of $\eta_C$ at non zero power in the irreversible region attracted much 
interest in recent years \cite{nonzero}.  The high values of efficiency obtained by 
the practical heat engines are not necessarily in the region of maximum power 
output \cite{holubec1}. Hence this model requires further study of the optimal 
efficiency  at arbitrary power \cite{HoluStat}.

It should be noted that the universal relation (\ref{maxeffuni1}) at maximum
power is valid not only for heat engine models having strong coupling with 
left-right symmetry but also for models (Curzon- Ahlborn heat engine and the Feynman ratchet) 
without having such symmetry \cite{Tu}. The previous study showed that the energy-matching condition 
is sufficient for obtaining  Eq.(\ref{maxeffuni1}) for different heat engine models \cite{Tu}. 
Obtaining such an energy matching condition for our generalized model 
will be a part of our future work.

\section*{Acknowledgment:} I would like to thank the anonymous referees for their critical comments
and supportive suggestions.

\end{document}